\documentclass[aps,preprint,showkeys,showpacs]{revtex4}
\usepackage{graphicx}
\begin{document}

\title{Coulomb dissociation of ${^9}$Li and the rate of the ${^8}$Li(n,$\gamma$)${^9}$Li reaction}
%
\author{P. Banerjee$^{1,2}$\footnote{Present address:
Department of Physics, Presidency College, 86/1 College Street, Kolkata -700 073, India}}
\email{banprabir@gmail.com}
\author{R. Chatterjee$^{1,3}$\footnote{Present address:
Department of Physics, Indian Institute of Technology, Roorkee -247 667, India}}
\email{rajdeep.chatterjee@ulb.ac.be}
\author{R. Shyam$^{1}$}
\email{radhey.shyam@saha.ac.in}
\affiliation{$^1$Theory Group, Saha Institute of Nuclear Physics, Kolkata -
700 064, India \\
$^2$Jhargram Raj College, Jhargram - 721 507, India \\
$^3$Physique Nucl\'eaire Th\'eorique et Physique Math\'ematique CP229, 
Universit\'e Libre de Bruxelles, B-1050 Brussels, Belgium
}

\begin{abstract}
We calculate the Coulomb dissociation of $^9$Li on Pb and U targets at 28.5 
MeV/A beam energy within a finite range distorted wave Born approximation
formalism of the breakup reactions. Invoking the principle of detailed balance,
these cross sections are used to determine the excitation function and 
subsequently the rate of the radiative capture reaction 
$^8$Li(n,$\gamma$)$^9$Li at astrophysical energies. Our method is free from
the uncertainties associated with the multipole strength distributions of the
$^9$Li nucleus. The rate of this reaction at a temperature of 10$^9$K is found
to be about 2900 cm$^3$ mole$^{-1}$ s$^{-1}$.  
\end{abstract}
\maketitle
\newpage

\section{Introduction}

The ${^8}$Li(n,$\gamma$)${^9}$Li reaction plays an important role 
in determining the amount of matter that can be produced at mass number
$A>$ 8. Inhomogeneous big bang nucleosynthesis and Type II supernova
are the proposed sites for such synthesis processes. In the first site,
after the production of $^7$Li the path to $A>$ 12 nuclei goes through the 
chain $^7$Li($n,\gamma)^8$Li$(\alpha,n)^{11}$B, with a weaker branch going 
through the $^7$Li($\alpha,\gamma)^{11}$B path (see, {\it e.g.}, 
Ref.~\cite{boy92,gu95}). However, the neutron capture on $^8$Li provides a leak
from this primary chain and depending on the rate of this reaction the 
production of nuclei with $A>$ 12 can reduce by 40-50$\%$~\cite{mal88}.

In the post-collapse phase of a type II supernova comes the opportunity to
produce heavy isotopes via the $r$-process. In the early expanding phase, 
starting with a He-rich environment the mass-8 gap would be bridged by either
$\alpha + \alpha + \alpha \to ^{12}$C or $\alpha + \alpha + n \to ^{9}$Be
reactions. These reactions would continue until a neutron rich freeze out
occurs which triggers the $r$-process~\cite{woo94}. At this stage it would
also be possible to bridge the A = 8 gap through the reaction chain 
$^4$He$(2n,\gamma)^6$He$(2n,\gamma)^8$He($\beta^-$)$^8$Li($n,\gamma)^9$Li
$(\beta^-)^9$Be~\cite{gor95,efr96}. This chain would provide an alternative 
path to proceed along the neutron-rich side of the line of stability towards 
heavier isotopes such as $^{36}$S, $^{40}$Ar, $^{46}$Ca, and $^{48}$Ca.
The origin of these neutron rich isotopes is under debate. It is of 
critical importance to know to what extent this chain competes with the 
$^8$Li($\beta^-$)$^8$Be($2\alpha$) process. An important clue to the answer
depends on knowing as accurately as possible the rate of the
$^8$Li($n,\gamma)^9$Li reaction and the neutron density. Reaction
chains similar to ones that are supposed to occur in type II supernova
can be found in the material ejected from neutron star mergers~\cite{ros01}
and thus the importance of the accurate knowledge of this reaction is 
emphasized again.

The rate ($R$) of a nuclear reaction where two nuclei $b$ and $c$ interact to
form the reaction products of the final channel is given by~\cite{rol88}
\begin{eqnarray}
{\rm R}  =  N_b N_c \langle \sigma(v_{bc}) v_{bc} \rangle 
(1+\delta_{bc})^{-1},
\label{r1}
\end{eqnarray}
where $N_b$ and $N_c$ represents the total number of nuclei $b$ and $c$ taking
part in the reaction and $\delta_{bc}$ is the Kronecker delta which is unity 
if $b$ and $c$ are identical and zero otherwise. $\sigma(v_{bc})$ is the 
cross section for a single target nucleus at the relative velocity of $v_{bc}$.
The number densities $N_i$ are related to the matter density $\rho$ and mole 
fraction $Y_i$ by $N_i = \rho N_A Y_i$, where $N_A$ is the Avogadro number. 
In Eq.~(\ref{r1}) the product $\sigma(v_{bc}) v_{bc}$ is averaged over the 
Maxwell- Boltzmann velocity distribution and is interpreted as the reaction 
rate per particle pair. This is given by
\begin{equation}
\langle \sigma(v_{bc})v_{bc} \rangle = \Big({8\over{(k_BT)^3}\pi\mu}\Big)^{1/2}
\int_0^{\infty} \sigma(E_{bc}) E_{bc} e^{-{E_{bc}/(k_BT)}} dE_{bc},
\end{equation}
where $\mu$ is the reduced mass of the interacting nuclei, $k_B$ is the
Boltzmann constant and $T$ is the relevant stellar temperature. $E_{bc}$ 
represents the energy corresponding to the relative velocity $v_{bc}$. 

It is thus clear from Eqs. (1)-(2) that knowledge of the reaction cross section
$\sigma(E_{bc})$ as a function of the relative velocity (or energy) in the 
astrophysically relevant energy region is the prime requirement for calculating
the rate of a particular reaction. 

Since big-bang nucleosynthesis starts when the temperature has fallen to
about 100 keV, the rate of the ${^8}$Li(n,$\gamma$)${^9}$Li reaction is of
astrophysical importance for neutron energies in similar range. This reaction
can proceed by both direct capture as well as via resonant capture through 
the 5/2$^-$ state of $^9$Li at the excitation energy of 4.296 MeV. For the 
inverse reaction $(\gamma,n)$, the resonant capture via this state would imply
the dominance of the E2 transition multipolarity in its excitation from the 
$3/2^-$ ground state of $^9$Li. The ratio of E2 to E1 excitation has been 
estimated in Ref.~\cite{ber99} for a $^9$Li projectile on a $^{208}$Pb target
at the beam energy of 28.5 MeV/nucleon. The maximum value of this ratio even
at the resonance peak (corresponding to $E_\gamma$ = 0.26 MeV) is only 0.018.
Within the energy range $E_\gamma$ = 0-1 MeV, this ratio is about 0.0023. 
Therefore, for the reaction ${^8}$Li(n,$\gamma$)${^9}$Li only the direct 
capture mechanism through E1 transition applies in this energy regime.

Several theoretical predictions of the rate of the $^8$Li($n,\gamma)^9$Li
have been reported. Some of them perform the nuclear structure calculations
of $^9$Li and calculate the capture cross sections from the corresponding
wave functions~\cite{mao91,des93}. Others estimate the rate of this reaction
from the systematics that are based on information existing for other 
nuclei~\cite{rau94,mal89}. These rates vary from each other by more than 
an order of magnitude. Hence, efforts have also been made to determine the 
rate of this reaction by experimental methods~\cite{zec98,kob03}. 

Since $^8$Li has a very small half-life ($\approx$ 838 ms), a direct 
measurement of the cross section ($\sigma_{n\gamma}^{^9Li}$) of the reaction 
$^8$Li(n,$\gamma$)$^9$Li is nearly impossible. However, with a beam of $^9$Li,
it is possible to measure the cross section ($\sigma_{\gamma n}^{^9Li}$) 
of the reverse reaction $^9$Li + $\gamma \to$ $^8$Li + $n$ 
(photodisintegration process), and use the principle of detailed balance to 
deduce the cross section $\sigma_{n\gamma}^{^9Li}$ as 
\begin{eqnarray}
\sigma_{n\gamma}^{^9Li} = 0.8 \frac{k_\gamma^2}{k^2}\sigma_{\gamma n}^{^9Li}
\end{eqnarray}
In Eq.~(3), the photon wave number is given by $k_\gamma = 
\frac{E_\gamma}{\hbar c} = \frac{(E_{n-{^8}{\rm Li}} + Q)}{\hbar c}$, 
in terms of the Q-value of the capture reaction with $E_{n-{^8}{\rm Li}}$  
being the center of mass (c.m.) energy of the $n-{^8}{\rm Li}$ system. $k$ 
is the wave vector corresponding to $E_{n-{^8}{\rm Li}}$.

A very promising way of studying the photodisintegration process
is provided by the virtual photons acting on a fast charged nuclear 
projectile when passing through the Coulomb field of a heavy target nucleus
\cite{bau86,bau03,bau08}. The advantage of this Coulomb dissociation (CD) 
method is that here measurements can be performed at higher beam energies 
which enhances the cross sections considerably as compared to the those of
the direct method. At higher energies the fragments in the final channel 
emerge with larger velocities which facilitates their more accurate detection.
Furthermore, the choice of the adequate kinematical condition of the 
coincidence measurements allows the study of low relative energies of the 
final state fragments and ensures that the target nucleus remains in the 
ground state during the reaction. However, the success of this method depends
on nuclear breakup effects being either negligible or at least their magnitude
being known as accurately as possible.
 
In the recent past attempts have been made to measure the CD of $^9$Li on U 
and Pb targets at the beam energy $(E_{beam})$ of 28.5 MeV/A \cite{zec98} and
on a Pb target at $E_{beam}$ of 39.7 MeV/A \cite{kob03}. The corresponding 
cross sections were used to get the photoabsorption cross sections 
$\sigma_{\gamma n}^{^9Li}$ by following the method of virtual photon number
\cite{bau86} 
\begin{eqnarray}
\sigma_{\gamma n}^{^9Li} & = & \frac{E_\gamma}{n_{E\lambda}}
\frac{d\sigma}{dE_\gamma},
\end{eqnarray}
where $n_{E\lambda}$ is the virtual photon number~\cite{ald66} of electric
multipole order $\lambda$ and $\frac{d\sigma}{dE_\gamma}$ is the measured CD
cross section. $\sigma_{n \gamma}^{^9Li}$ can be obtained from 
$\sigma_{\gamma n}^{^9Li}$ by using Eq.~(3) which can be used to get the rate
of the reaction $^8$Li(n,$\gamma$)$^9$Li. In Ref.~\cite{zec98} $R$ was 
estimated to be $<$ 7200 cm$^3$ s$^{-1}$ mole$^{-1}$, while it was 
reported to be $<$ 790 cm$^3$ s$^{-1}$ mole$^{-1}$ in Ref.~\cite{kob03}.  

\section{Method of Calculation}
In the theoretical determination of $\sigma_{n\gamma}^{^9Li}$ within the CD
approach, one calculates the Coulomb dissociation cross sections of $^9$Li
by using a theory of the CD process. In Ref.~\cite{ber99}, first order 
Coulomb excitation theory has been used for this purpose. A crucial quantity 
that enters in calculations within this theory is the reduced transition 
probability $B(E\lambda$) of a particular transition. This quantity depends
on the wave function of the relative motion of $n$ and $^8$Li in the ground as 
well as excited states of $^9$Li. Since CD method involves excitation of 
the projectile to its continuum, the evaluation of $B(E\lambda)$ depends 
sensitively on information about the continuum structure of the projectile.
In Ref.~\cite{ber99} the continuum states were calculated by treating them
as scattering states with the same $n-^8$Li potential which was obtained by
fitting the binding energy of the $^9$Li ground state. The calculated CD
cross section is used to get $\sigma_{n\gamma}^{^9Li}$ with help of Eq.~(3).
From the comparison with the experimental capture cross sections 
$\sigma_{n\gamma}^{^9Li}$ of Ref.~\cite{zec98} a value of $<$ 2200 cm$^3$ 
s$^{-1}$ mole$^{-1}$ have been obtained for the rate of the 
$^8$Li(n,$\gamma$)$^9$Li reaction. We note that this value differs from that 
reported in Ref~\cite{zec98} by a factor of about 4. 

In this paper we use a fully quantum mechanical theory of Coulomb breakup 
reactions to calculate the Coulomb dissociation of ${^9}$Li which is then
used to extract the rate of the capture reaction ${^8}$Li(n,$\gamma$)${^9}$Li.
The theory of CD reactions used by us is formulated within the post form 
finite range distorted wave Born approximation (FRDWBA)~\cite{cha00} where 
the electromagnetic interaction between the fragments and the target nucleus
is included to all orders and the breakup contributions from the entire 
non-resonant continuum corresponding to all the multipoles and the relative 
orbital angular momenta between the fragments are taken into account
\cite{pra02}. Full ground state wave function of the projectile, of any 
orbital angular momentum configuration, enters as an input into this theory. 
Unlike the theoretical models used in Ref.~\cite{ber99}, this model does not 
require the knowledge of the positions and widths of the continuum states. 
Thus our method is free from the uncertainties associated with the multipole 
strength distributions occurring in other formalisms as we need only the 
ground state wave function of the projectile as input.

Let us consider the reaction $ a + t \rightarrow b + c + t $, where the 
projectile $a$ breaks up into fragments $b$ (charged) and $c$ (uncharged) in
the Coulomb field of a target $t$. The relative energy spectra for the 
reaction is given by 
\begin{eqnarray}
{d\sigma \over dE_{bc} }=\int_{\Omega _{bc},\Omega_{at}} d\Omega _{bc} 
d\Omega_{at} \left\{\sum_{l m}\frac{1}{(2l + 1)}\vert \beta_{lm}\vert^2 
\right\} {2\pi\over \hbar v_{at}} {{\mu_{bc}\mu_{at}p_{bc}p_{at}} \over 
{h^6}}~ \label{cs},
\end{eqnarray}
where $v_{at}$ is the $a$--$t$ relative velocity in the entrance channel, 
$\Omega _{bc}$ and $\Omega_{at}$ are solid angles, $\mu_{bc}$ and $\mu_{at}$
are reduced masses, and $p_{bc}$ and $p_{at}$ are appropriate linear momenta 
corresponding to the $b$--$c$ and $a$--$t$ systems, respectively. 
 
The reduced amplitude $\beta_{lm}$ in the post form finite range distorted 
wave Born approximation is given by
\begin{eqnarray}
\beta_{lm}
= \langle \exp(\gamma\vec{k}_c-\alpha\vec{K})\vert V_{bc}\vert 
\Phi _{a}^{lm}\rangle \langle \chi ^{(-)}(\vec{k}_b)\chi^{(-)}
(\delta\vec{k}_c)\vert \chi^{(+)}(\vec{k}_a) \rangle~, \label{dw}
\end{eqnarray}
where, $\vec{k}_b$, $\vec{k}_c$ are Jacobi wave vectors of fragments $b$
and $c$, respectively in the final channel of the reaction, $\vec{k}_a$ is 
the wave vector of projectile $a$ in the initial channel and $V_{bc}$ is the 
interaction between $b$ and $c$. $\Phi _{a}^{lm}$ is the ground state wave 
function of the projectile with relative orbital angular momentum state $l$ 
and projection $m$. In the above, $\vec{K}$ is an effective local momentum 
associated with the core-target relative system, whose direction has been 
taken to be the same as the direction of the asymptotic momentum $\vec{k}_b$ 
\cite{shy85,cha00}. $\alpha, \delta$ and $\gamma$ in Eq. (5), are mass factors 
relevant to the Jacobi coordinates of the three body system (see Fig. 1 of 
Ref. \cite{cha00}). $\chi^{(-)}$'s are the distorted waves for relative motions
of $b$ and $c$ with respect to $t$ and the c.m. of the $b-t$
system, respectively, with ingoing wave boundary condition and 
$\chi^{(+)}(\vec{k}_a)$ is the distorted wave for the scattering of the c.m. 
of projectile $a$ with respect to the target with outgoing wave boundary 
condition.

It should be mentioned that in Eq. (\ref{dw}), the interactions between the
fragments $b$ and $c$ and the target nucleus are included to all orders, but
the $b$-$c$ interaction is treated to first order only. Since for relative
energies of our interest there are no resonances in the $n$+$^8$Li continuum,
we expect this approximation to be valid. It is clearly a good approximation 
for the deuteron and the neutron halo systems~\cite{bau03}. For those cases 
where higher order effects of the fragment-fragment interaction are known to 
be non-negligible, our model will have a limited applicability. It should
be noted that in calculating the relative energy spectra within this theory 
explicit information about the continuum strength distribution of the 
projectile is not required; the entire continuum is automatically included in 
our post form theory.
 
Physically, the first term in Eq. (\ref{dw}) contains the structure information
about the projectile through the ground state wave function $\Phi _{a}^{lm}$,
and is known as the vertex function, while the second term is associated only 
with the dynamics of the reaction. The charged projectile $a$ and the fragment
$b$ interacts with the target by a point Coulomb interaction and hence 
$\chi^{(-)}_b(\vec{k}_b)$ and $\chi^{(+)}(\vec{k}_a)$ are substituted with
appropriate Coulomb distorted waves. For pure Coulomb breakup, of course, the 
interaction between the target and the uncharged fragment $c$ is zero and hence
$\chi^{(-)}(\delta\vec{k}_c)$ is replaced by a plane wave. This will allow the
second term of Eq. (\ref{dw}) to be evaluated analytically in terms of the 
bremsstrahlung integral \cite{nor54}. A more detailed description of how the
reduced amplitude $\beta_{lm}$  is simplified and analytical expression for 
the bremsstrahlung integral, as used in our case, can be found in Refs. 
\cite{cha00,pra02}.  

One can then relate the cross section in Eq. (\ref{cs}) to the 
photodissociation cross section, $\sigma_{\gamma n}^a$, for the reaction 
$a + \gamma \rightarrow b + c$, by
\begin{equation}
{d\sigma \over {dE_{bc}}} = {1\over {E_{\gamma}}}\sum_{\lambda} n_{\pi\lambda}
\sigma_{\gamma n}^a, \label{ne}
\end{equation}
where  $n_{\pi\lambda}$ is the equivalent photon number of type $\pi$ (
electric or magnetic) and multipolarity $\lambda$~\cite{bau86}, and 
photon energy $E_{\gamma} = E_{bc} + S_n$, with $S_n$ being the one 
neutron separation energy of the projectile $a$. The relative energy between 
the fragments in the final state is denoted by $E_{bc}$. As discussed 
earlier, for the case of our interest, transition of E1 multipolarity
dominates. The virtual photon number for this case has been calculated by
following the same method as that used in Ref.~\cite{ber99}.

Of course, the procedure of relating the CD cross sections calculated by Eq.
(\ref{cs}) to $\sigma_{\gamma n}^a$ by Eq.~(7) is valid only when transitions 
of a single multipolarity and type give the dominant contribution to the 
breakup cross section and nuclear breakup effects are negligible. Both of 
these conditions are supposed to be fulfilled in the case of our interest. 
Furthermore, the post-form amplitudes [Eq. (\ref {cs})] include 
fragment-target interaction to all orders while the right hand side of
Eq.~(7) has been written within first order perturbation theory. Therefore,
relating the post-form CD cross section to the photodissociation cross 
sections via Eq.~(7) is valid only when higher order effects make negligible 
contribution to the CD cross sections at the relevant beam energy. Indeed,
it has been shown in Ref.~\cite{pra02} that for the Coulomb breakup reaction 
involving projectiles of similar mass range, the higher order effects are
almost negligible for $E_{bc}$ of our interest ($<$ 100 keV) for beam energies
around 30 MeV/nucleon. Therefore, necessary conditions for the validity of 
Eq.~(7) are fulfilled for the present case. Nevertheless, we emphasize that 
in general, the validity of Eq.~(7) in each case must be checked before using
this to extract the photodissociation cross section from the post-form Coulomb
dissociation cross section.

The radiative capture cross section, $\sigma_{n\gamma}^a$, for the reaction, 
$b + c \rightarrow a + \gamma $, is then related to the photo dissociation 
cross section, $\sigma_{\gamma n}^a$, by the principle of detailed balance 
\cite{bau86} (Eq.~(3) for the reaction of our interest).   

\section{Results and discussions}
As shown above, the ground state wave function of the projectile enters into
the calculations of the CD cross sections within our theory.  For $^{9}$Li, 
we obtain this wave function by assuming the neutron-$^8$Li core 
interaction to be of Woods-Saxon type whose depth is searched, for a given 
configuration, to reproduce the corresponding binding energy.

Within this model, the valence neutron in $^9$Li (spin-parity $3/2^-$) is 
assumed to move relative to an inert $^8$Li core (with intrinsic spin-parity 
$2^+$) in a Woods-Saxon plus spin-orbit potential, with an adjustable depth 
$V_0$ for the initial channel: 
\begin{eqnarray}
V(r)=V_0(1-F_{s.o.}({\vec l}.{\vec s}){r_0\over r}
{d\over dr})f(r),
\end{eqnarray}
where
\begin{eqnarray}
f(r)=(1+\exp((r-R)/a))^{-1}.
\end{eqnarray}
Using $a$=0.52 fm, $r_0$=1.25 fm, $^8$Li core radius, $R$=2.49 fm and the 
spin-orbit strength, $F_{s.o.}$ = 0.351 fm, the depth of the Woods-Saxon 
potential was searched to reproduce the one-neutron separation energy of 
$^9$Li (4.05 MeV). This yielded $V_0 = -45.21$ MeV. With this potential, 
the rms distance of the core-neutron relative motion and the rms size of 
$^9$Li came out to be 3.10 fm and 2.55 fm, respectively.

\begin{figure}[ht]
\centering
\includegraphics[width=.40\textwidth,clip=]{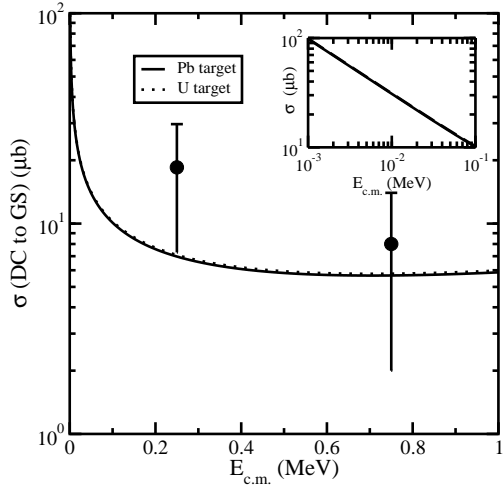}
\caption{\label{fig1} 
Direct capture (DC) cross sections to the ground state (GS) of $^{9}$Li.
The solid and dotted curves (which almost coincide with each other)
are calculated using the Coulomb dissociation of $^9$Li on Pb and U targets
at 28.5 MeV/A beam energy. The inset shows the values of the capture 
cross sections upto ${\rm E_{c.m.}}$ $\leq$ 100 keV.
The experimental data are from Ref. \cite{zec98}.} 
\end{figure}

We have calculated the capture cross sections of the $^8$Li(n,$\gamma$)$^9$Li
reaction as a function of the c.m. relative energy (${\rm E_{c.m.}}$)
between neutron and $^8$Li in the range of (0 - 1) MeV, using the Coulomb 
breakup cross section obtained in our method. Since, our aim in this paper,
is to narrow down the theory dependent uncertainty in the extracted rate of 
this reaction, we compare our results with the same experimental capture cross
sections as those in Ref.~\cite{ber99}. In Fig. 1, we show the direct capture 
cross sections to the ground state of $^{9}$Li obtained from the Coulomb 
dissociation of $^{9}$Li on Pb (solid line) and U targets (dotted line) at 
28.5 MeV/A beam energy. In the inset of this figure we have highlighted the
values of the cross sections in the astrophysically interesting region 
(for ${\rm E_{c.m.}}$ $\leq$ 100 keV ) by presenting cross sections as a 
function of ${\rm E_{c.m.}}$ on a log-log plot. As expected, the capture cross
section is independent of the target used during the Coulomb dissociation. It 
should be noted that while we have used a spectroscopic factor ($S$) of 
$0.68 \pm 0.14$ for the ground state of $^9$Li which has been extracted 
recently from a transfer reaction measurement~\cite{li005}, a shell model 
value of 0.94 was used for it in Ref.~\cite{ber99}. It is worth mentioning 
that transfer reaction cross sections are very sensitive to the angular 
momentum state of the projectile and hence have been widely used to extract 
nuclear spectroscopic factors. Had we used the shell model value of $S$, our 
results would have been proportionately higher.
 
Nevertheless, it should be noted that the experimental data of 
Ref.~\cite{zec98} have uncertainty of approximately a factor of 2. Furthermore,
the second Coulomb dissociation measurement of $^9$Li as reported in
Ref.~\cite{kob03} indicates that the extracted capture cross section could
even be substantially lower than those reported in Ref.~\cite{zec98}. 
Therefore, to firm up the theoretical capture cross sections as extracted
from the Coulomb dissociation method, the uncertainty in the experimental
data should be minimized as much as possible. 

\begin{figure}
\centering
\includegraphics[width=.40\textwidth,clip=]{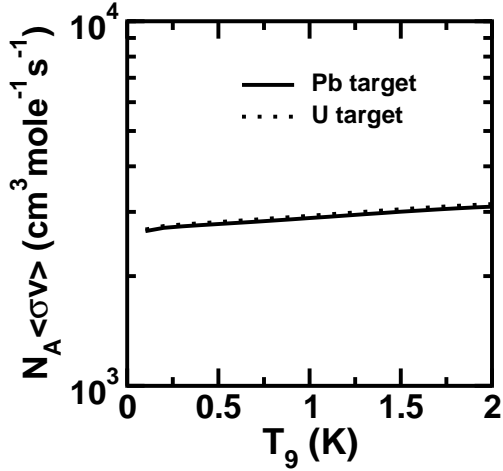}
\caption{\label{fig2} 
Capture rates for the ${^8}$Li(n,$\gamma$)${^9}$Li reaction as a function of
temperature in units of 10$^9$K. Solid and dotted lines are reaction rates 
derived from the Coulomb dissociation of ${^9}$Li on Pb and U targets, 
respectively.}
\end{figure}

Reaction rate ($R$) calculated from the capture cross sections are 
plotted in Fig.~2 as a function of T$_9$ (the temperature equivalent of 
relative energy in units of 10$^9$K). Solid and dotted lines show reaction 
rates derived from the Coulomb dissociation of ${^9}$Li on Pb and U targets, 
respectively. The rate changes in the range (2800 - 3100) cm$^3 $mole$^{-1}$ 
s$^{-1}$ for T$_9$ between 0.5 and 2 and the value at T$_9$ = 1 is 
approximately 2900 cm$^3$mole$^{-1}$ s$^{-1}$, when averaged over the two 
targets.  

As is evident from the integrand in Eq.~(2), for a fixed stellar temperature, 
the maximum contribution to the reaction rate is highly dependent on the 
reaction cross section and in turn on the relative energy. At T$_9$ = 1, 
the maximum contribution to the ${^8}$Li(n,$\gamma$)${^9}$Li reaction rate 
comes from a low relative energy of 45 keV. At this low energy it is extremely
difficult to measure reaction cross sections by direct methods. This
is where the power of the CD method becomes more evident as an indirect method
in nuclear astrophysics. With recent advances in experimental techniques 
it is possible to measure relative energy spectra at quite low relative 
energies. 
\begin{table}
\caption{The comparison of reaction rates of the ${^8}$Li(n,$\gamma$)${^9}$Li 
reported as reported by various authors} 
\begin{ruledtabular}
\begin{tabular}{ll}
\hline
 Reference & Reaction rate $(cm^3 mol^{-1} s^{-1})$  \\
\hline
Malaney and Fowler~\protect\cite{mal89} & 43000   \\
Mao and Champagne~\protect\cite{mao91} & 25000  \\
Descouvemont~\protect\cite{des93} & 5300  \\
Rauscher {\it et al.}~\protect\cite{rau94}  & 4500  \\
Zecher {\it et al.}~\protect\cite{zec98}    & $<$ 7200 \\
Kobayashi {\it et al.}~\protect\cite{kob03} & $<$ 790  \\
Bertulani~\protect\cite{ber99} & 2200 \\
Present work                   & 2900
\end{tabular}
\end{ruledtabular}
\end{table}

In Table I, we present a comparison of the rates of the reaction
${^8}$Li(n,$\gamma$)${^9}$Li reported by various workers. It is interesting
to note that the rate of the ${^8}$Li(n,$\gamma$)${^9}$Li reaction extracted 
by us is  within 30$\%$ in agreement with that computed from the capture cross
sections of Ref.~\cite{ber99} where a completely different theoretical model of
CD process was used. On the other hand, our rate is about 45-35$\%$ smaller than
those reported in Refs.~\cite{des93,rau94} where they have been obtained from
structure model calculations of $^9$Li. Our values are in  sharp disagreement 
with the results of Ref.\cite{mao91} where calculations were performed 
within the spd-shell model and with those of Ref.\cite{mal89} which have been
obtained from the systematics of similar nuclei. The rate of Ref.~\cite{mao91}
is larger by a factor of 7.2 whereas that of Ref.~\cite{mal89} is even larger
(by a factor of almost 15). It may be worthwhile to see what these calculations 
would predict if the latest experimental information on the spectroscopic 
factor for the $^9$Li $\rightarrow ^8$Li + $n$ partition was taken into
consideration. 

Thus, our calculations do not support the large rate for the 
${^8}$Li(n,$\gamma$)${^9}$Li reaction. This would suggest that a significant 
portion of the ${^8}$Li would remain available for alpha-capture to $^{11}$B 
and would not be destroyed by the ${^8}$Li(n,$\gamma$)${^9}$Li reaction.  
Therefore, the ${^8}$Li(n,$\gamma$)${^9}$Li reaction does not hamper the 
formation of $A>$12 elements through the 
$^8$Li($\alpha$,n)$^{11}$B(n,$\gamma$)$^{12}$Be$(\beta^-)^{12}$C$(n,\gamma)..$
reaction chain.

\section{Summary and Conclusions}
In summary, we have calculated the rate of the ${^8}$Li(n,$\gamma$)${^9}$Li
reaction by studying the inverse photodissociation reaction in terms of the
Coulomb dissociation of $^9$Li on heavy targets at 28.5 MeV/A using a theory
formulated within the finite range post form distorted wave Born 
approximation. This capture reaction provides, in an inhomogeneous early 
universe, a leak from the primary chain of nucleosynthesis, thereby reducing
the production of heavy elements. The advantage of our theoretical method is 
that it is free from the uncertainties associated with the multipole strength 
distributions of the projectile. The newly extracted experimental ground state
spectroscopic factor for the  ${^9}$Li $\to $ $^8$Li $+ n$ 
partition~\cite{li005}, has been incorporated in our theory.

The rate of this reaction at a temperature of 10$^9$K has been found to be 
about 2900 cm$^3$ mole$^{-1}$ s$^{-1}$. This value is in agreement (within 
30$\%$) with the earlier Coulomb dissociation analysis of this data using a
different theoretical model. Thus theoretical uncertainty in the rate of
${^8}$Li(n,$\gamma$)${^9}$Li reaction as determined from the Coulomb 
dissociation of $^9$Li is much lower than the experimental
uncertainties in this data. Therefore, it would be worthwhile to make more
precise measurements of the Coulomb dissociation reaction. The maximum 
contribution to the 
reaction rate at this stellar temperature, came from a low relative 
energy of 45 keV. Thus in future experiments an attempt should be made 
to measure the ${^8}$Li(n,$\gamma$)${^9}$Li capture cross section at this 
low relative energy to get a more accurate picture of the reaction rate.

Our calculations also suggest that this reaction rate is not high enough 
to destroy enough of $^8$Li so as to significantly reduce the formation of 
$A>$12 elements through the
$^8$Li($\alpha$,n)$^{11}$B(n,$\gamma$)$^{12}$Be$(\beta^-)^{12}$C$(n,\gamma)..$ reaction chain. 

\section{Acknowledgments}
PB and RC wish to thank the Theory Group of Saha Institute
of Nuclear Physics for their hospitality during this collaboration.
 
\end{document}